\documentclass[letterpaper]{article}

\usepackage[T1]{fontenc}
\usepackage{hyperref}
\usepackage{geometry}
\usepackage{setspace}

\usepackage[style = chem-acs, doi = true, articletitle=true]{biblatex}
\usepackage{graphicx}
\usepackage{float}
\newfloat{scheme}{htbp}{los}
\floatname{scheme}{Scheme}
\floatname{chart}{Chart}
\newfloat{graph}{htbp}{loh}

\usepackage{chemformula} 
\usepackage[version = 4]{mhchem} 

\newcommand{\etal}{\textit{et al}.\ }

\newcommand{\fcunit}{eV/\AA$^2$\ }

\usepackage{authblk}
\author[1]{Mohammad Bagheri*}
\affil[1]{Nanoscience Center, Department of Physics, University of Jyv\"askyl\"a, 40014 Jyv\"askyl\"a, Finland}
\author[2]{Ethan Berger}
\affil[2]{Department of Physics, Chalmers University of Technology, SE-41296 Gothenburg, Sweden}
\author[3]{Hannu-Pekka Komsa}
\affil[3]{Microelectronics Research Unit, Faculty of Information Technology and Electrical Engineering, University of Oulu, P.O. Box 4500, Oulu, FIN-90014, Finland}
\author[1]{Pekka Koskinen*}

\title{Massive Discovery of Low-Dimensional Materials \\from Universal Computational Strategy}
\date{*Email: mohammad.m.bagheri@jyu.fi, pekka.j.koskinen@jyu.fi}

\begin{document}

\maketitle

\begin{abstract}
  Low-dimensional materials have attractive properties that drive intense efforts for novel materials discovery. However, experiments are tedious for systematic discovery, and present computational methods are often tuned to two-dimensional (2D) materials, overlooking other low-dimensional materials. Here, we combined universal machine-learning interatomic potentials (UMLIPs) and an advanced, interatomic force constant (FC) -based dimensionality classification method to make a massive discovery of novel low-dimensional materials. We first benchmarked UMLIPs' first-principles-level accuracy in quantifying FCs and calculated phonons for $35{,}689$ materials from the Materials Project database. We then used the FC-based method for dimensionality classification to discover $9139$ low-dimensional materials, including $1838$ 0D clusters, $1760$ 1D chains, $3057$ 2D sheets/layers, and $2484$ mixed-dimensionality materials, all of which conventional geometric descriptors have not recognized. By calculating the binding energies for the discovered 2D materials, we also identified $887$ sheets that could be easily or potentially exfoliated from their parent bulk structures.
\end{abstract}





\section{Introduction}

The discovery of novel functional materials for targeted applications has long been central in materials research. Over the past two decades, the rise of graphene has revealed remarkable properties that have sparked a quest for novel two-dimensional (2D) materials with ever-growing attention  \cite{Novoselov2004}. To date, the quest has unearthed numerous 2D materials such as transition-metal dichalcogenides (TMDs), MXenes, metallenes, and emerging non‑van der Waals systems, offering promising applications in electronics \cite{Radisavljevic2011, Butler2013, Manzeli2017, Xiang20_Sci, Han19_NComm}, optics \cite{Du21_NComm, Wei21_AM, Turunen22_NRevPhys}, magnetism \cite{Burch2018, Gibertini2019, Klein23_ACSNano}, sensors \cite{Kim_2018, mx_sens, Zhou_2021}, catalysis \cite{Deng2016, D4NR00147H, Zhang_2025, Xie2023}, and topological materials \cite{Kou2017, Balandin22_MT, Campi2021, Kezilebieke2020}.

Another approach to materials discovery is computation. Many studies have employed high-throughput computations and data-driven approaches to systematically screen thousands of experimentally known bulk materials, uncovering hundreds to over a thousand exfoliable 2D material candidates \cite{Bjorkman2012, PRX2013, Ashton2017, Mounet2018, Lyngby2022, Arkady2022, Barnowsky2023, barnowsky2025l}. Yet these efforts have been targeted at 2D materials, with only scant studies systematically addressing materials with zero-dimensional (0D), one-dimensional (1D), or mixed-dimensional characteristics \cite{Hadeel2022, Screening2022, Li2024, Khazaei_lowD}.

To identify materials' low-dimensional character, most computational methods use geometrical criteria, such as atomic positions and atomic radii \cite{MP_dimen, Larsen_2019, Cheon2017, Gorai2016}. This approach has often proven reasonable for dimensionality identification. Geometries are usually calculated by density-functional theory (DFT), which is the workhorse method accurate enough for predicting materials and their properties. Yet even a realistic geometry can be inaccurate for dimensionality prediction, because it relies on empirical correlations between the lengths and strengths of atomic bonds. This shortcoming was recently remedied by a method that uses interatomic force constants (FCs) to identify low-dimensional units from bulk materials \cite{FCDimen}. While this method---FCDimen---has the merit of relying on chemical bonding quantitatively, so far its applicability has been limited by the high cost of DFT-level calculation of FCs. 

Fortunately, universal machine-learning interatomic potentials (UMLIPs) have recently become increasingly helpful in reducing calculation time and generating results at DFT-level accuracy \cite{Deringer19_AM, Behler2021_review, Jinnouchi2025_review}.  UMLIPS are trained by massive datasets and aim for applicability to all materials that contain nearly all elements. Although trained to predict only energies and forces, UMLIPs have been shown to reliably predict also their derivatives, such as formation energies of bulk materials \cite{Riebesell2024matbench}, defect formation energies \cite{Burger2025}, and phonon dispersion curves \cite{alexandria}. It would be reasonable to expect that UMLIPs would also be applicable in predicting FCs for fast and reliable identification of low-dimensional units.

Therefore, in this article, we used UMLIPs to calculate FCs and FCDimen to determine the dimensionalities of $35{,}689$ materials from the Materials Project (MP) database \cite{mp}. To choose the most accurate model, we benchmarked two UMLIPs, MatterSim \cite{MatterSim} and MACE \cite{mace1,mace2}, against DFT-calculated phonon properties and FCs of an existing phonon database with more than $10$K materials \cite{Togo2015, alexandria}. We then compared the FCDimen-predicted dimensionalities that DFT and UMLIPs gave for these materials, and concluded that MatterSim is sufficiently accurate for predicting material dimensionalities. Leveraging on MatterSim, we screened $153{,}234$ bulk materials and discovered $9139$ low-dimensional materials, including 2D sheets, 1D chains, 0D clusters, and various mixed dimensionalities. Finally, by calculating the interlayer binding energies for the discovered 2D sheets, we found that $887$ of them---all new to the known 2D materials databases---are easily or potentially exfoliable from their parent bulk structure.

\section{Methods}
\subsection*{Computational methods}
The machine-learning (ML) calculations were performed using MatterSim with pre-trained model v1.0.0-5M \cite{MatterSim} and MACE \cite{mace1,mace2} with the pre-trained large model of MACE-MP-0 (MACE-MP-0b3) without dispersion.
The MatterSim-v1 model is based on the M3GNet \cite{m3gnet} architecture, which was trained on relaxations obtained from the Materials Project \cite{mp}.
Both UMLIP models are available as calculators in the Atomic Simulation Environment (ASE) \cite{ase-paper}. ASE was used, with the help of Python, to construct the calculation workflow.
In the benchmark, all parameters were kept the same as those in the Phonon database \cite{Togo2015}.
All materials were relaxed using the FIRE \cite{FIRE} and BFGS algorithm to forces below $1$~meV/\AA.
The crystal structures and their properties were extracted using MP-API \cite{MP_API} and Python materials genomics (Pymatgen) \cite{pymatgen}. The space groups are calculated using spglib \cite{spglib} with symmetry precision $10^{-6}$.

Initial dimensionalities were calculated using Larsen \cite{Larsen_2019} method, which is implemented in ASE \cite{ase-paper}.
Phonon calculations were done using the Phonopy package \cite{phonopy1, phonopy2}, using a displacement of $0.01$ \AA. The supercell sizes used for determining the force constants were chosen so that the number of atoms was less than or equal to $250$ while keeping the point groups of the Bravais lattices of the conventional unit cell and the supercell the same.

Initial structures of known 2D materials (Figure~\ref{fig:d3}a) were extracted from Ref.~\cite{Mounet2018}.
All density-functional theory calculations were performed using VASP (Vienna Ab initio Simulation Package) software \cite{Kresse1996, PhysRevB.54.11169}, together with the projector-augmented wave method \cite{PhysRevB.50.17953}. The Becke-Johnson (BJ) damping variant of the DFT-D3 functional was adopted \cite{Grimme_BJ}. The plane wave cutoff is set to $520$ eV. The Brillouin zone of the unit cell is sampled by a $\Gamma$-centered k-point mesh whose density is defined by $R_k=20$, which in VASP determines the subdivisions $N_1$, $N_2$, and $N_3$ along the reciprocal lattice vectors $b_1$, $b_2$, and $b_3$, respectively, via $N_i = {\rm max}(1, R_k |b_i| + 0.5)$ and rounded to an integer. The energy convergence criterion between consequential steps is set to $10^{-8}$ eV for the self-consistent loop and $10^{-6}$ eV for ionic relaxation. Structures were considered to be relaxed when no force exceeded $1$~meV/\AA.

In the machine-learning binding energy calculation, we employed torch-dftd3 \cite{Grimme2006, Grimme2010, Grimme_BJ, Takamoto2022} alongside MatterSim to add the D3 correction in our calculations to correct the weak van der Waals (vdW) interaction. This approach has already proven successful for the investigation of 2D material heterobilayers \cite{Kristian2025}.

\subsection*{Binding energy calculation}
The method used to obtain the binding energy $E_b$ is adapted from Ref.~\cite{Bjorkman2012}. Taking materials from MP, the bulk unit cell is first relaxed with D3 correction to obtain the initial energy $E_0$ as well as the force constant. Using the latter, 2D units are identified using FCDimen. The resulting monolayers are then isolated and relaxed, leading to the layer energies $E_l$. The binding energy $E_b$ is then found as
\begin{equation}
E_{b} = \frac{\sum E_{l} - E_0}{A\cdot N_l},
\label{exfo}
\end{equation}
where $N_l$ is the number of layers in the unit cell and $A$ is the area of the layers.

\section{Results}

\subsection{FCDimen}
We begin by briefly reviewing the FC-based dimensionality classification approach FCDimen previously introduced in Ref.\ \cite{FCDimen}.
For a system with potential energy $U(\textbf{r}_{i},...,\textbf{r}_{n})$,
where $n$ is the number of atoms and $\textbf{r}_{i}$ the position of atom $i$, the force on atom $i$ is
\begin{equation}
F_{i}^{\alpha} = - \frac{\partial U}{\partial \textbf{r}_{i}^{\alpha}},
\end{equation}
where $\alpha$ is the Cartesian coordinate index. The FC between the coordinates $\alpha$ and $\beta$ of atoms $i$ and $j$ is
\begin{equation}
\Phi_{ij}^{\alpha\beta} = \frac{\partial^2 U}{\partial \textbf{r}_{i}^{\alpha} \partial \textbf{r}_{j}^{\beta}} = - \frac{\partial F_{i}^{\alpha}}{\partial \textbf{r}_{j}^{\beta}}.
\end{equation}
A scalar magnitude of FC between atom pair $ij$ is given by reducing the 3$\times$3 tensor using the Frobenius norm
\begin{equation}
\Phi_{ij} = ||\Phi^{\alpha\beta}_{ij}||_F.
\end{equation}
The strongest bond for atom $i$ can be identified by defining the maximum FC for each atom as
\begin{equation}
  \Phi_{i}^{\rm max} = \max_{j\neq i} ( \Phi_{ij} ),
  \label{eq:phi}
\end{equation}
which allows us to quantify minimum and maximum bond strengths in the material as  
\begin{align}
  {\rm Min FC} &= \min_i ( \Phi_{i}^{\rm max} ), \; \text{and} \nonumber \\ 
  {\rm Max FC} &= \max_i ( \Phi_{i}^{\rm max} ). 
  \label{eq:minmax}
\end{align}

To identify low-dimensional units, we treat the atoms as nodes in a graph. A pair of atoms is considered connected when the interatomic FCs are larger than a threshold $t={\rm Min FC}$, i.e., when $\Phi_{ij} \geq t$. The low-dimensional units are then defined as the resulting connected components of atoms linked by bonds. The dimensionality of these components is determined by the periodic extension of their unit cell (e.g., a component connected to its own periodic image in two directions is classified as 2D).
Furthermore,  scanning the threshold between $t={\rm Min FC} \ldots {\rm Max FC}$ gives a score in the range 0--1 for the assigned dimensionality. The FCDimen has proven promising for identifying low-dimensional materials and revealing new non-van der Waals candidates (see Ref. \cite{FCDimen} for more details).

\subsection{Benchmarking}
Before using FCDimen to classify dimensionalities, we must first ensure the reliable calculation of FCs. Loew \etal have shown that UMLIPs can predict phonon dispersion curves accurately \cite{alexandria}, with MACE \cite{mace1,mace2} and MatterSim \cite{MatterSim} performing the best. We therefore adopted these two UMLIPs for benchmarking [see Methods for details]. Reference FCs were taken from the Phonon database \cite{Togo2015}, which includes phonon properties at the DFT level for $10{,}032$ structures. This database was initially built using the PBEsol exchange-correlation functional. However, because both machine-learning (ML) models were trained on data based on the PBE functional, a fair comparison required the use of Alexandria's version of the Phonon database that was recalculated using the PBE functional \cite{alexandria}. For clean data and better comparisons, we applied the same screening criteria to the database as Ref.~\cite{FCDimen}. Screening involved the following criteria: (i) enforcing a minimum supercell lattice constant of 5~\AA; (ii) requiring that FCs at half the lattice distance remain below 20\% of the MaxFC to ensure the supercell size was sufficient for force constant (FC) decay; and (iii) filtering out chemically inert noble gases with negligible forces. Finally, we restricted our dataset to dynamically stable structures by excluding any materials exhibiting imaginary phonon frequencies (no imaginary phonon frequencies $>0.1$~THz) as determined by either DFT or ML.
We compared only MaxFC and MinFC, not the entire FC matrix. 

\begin{figure}[!h]
\begin{center}
  \includegraphics[width=6 in]{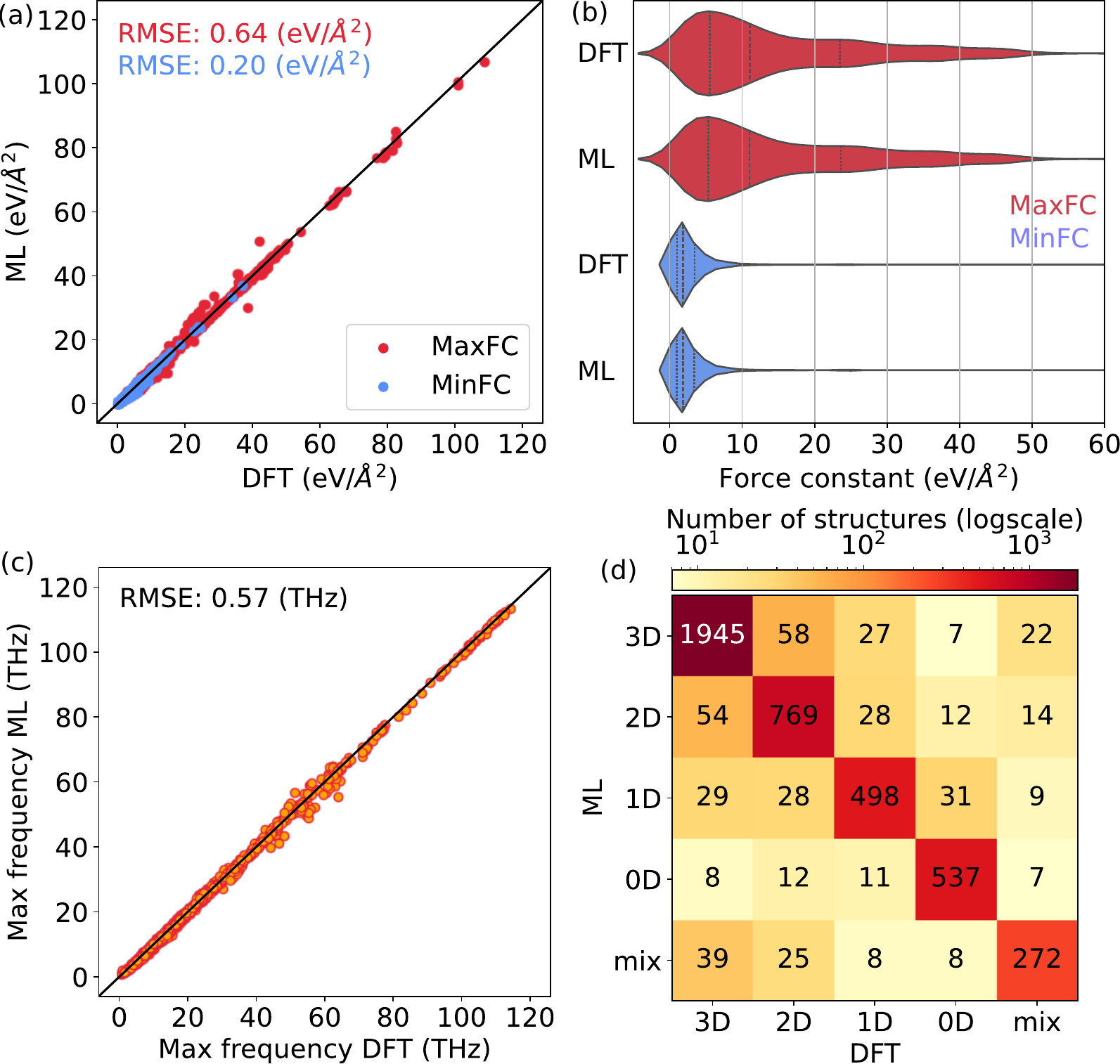}
\end{center}
\caption{\label{fig:benchmark}
Validation of methodology.
Comparing DFT and ML (MatterSim) for: a) MaxFC and MinFC [Equation~(\ref{eq:minmax})], b) distribution of MaxFC and MinFC, c) maximum phonon frequencies at $\Gamma$-point, and d) number of Phonon database structures in each dimensionality group. Panels (a) and (c) also show the related RMSE.
}
\end{figure}

MatterSim predicted FCs accurately, with a root mean squared error (RMSE) of $0.64$~\fcunit for MaxFC and $0.2$~\fcunit for MinFC (Figure~\ref{fig:benchmark}a). MACE was slightly less accurate, with RMSEs of $0.86$ \fcunit and $0.34$ \fcunit (Supplementary Information (SI) Figure~S1a). The FC distributions for ML and DFT are very similar (Figure~\ref{fig:benchmark}b and SI Figure~S1b). We remark that the RMSEs between PBE and PBESol were even larger, $0.46$~\fcunit for MinFC and $0.99$~\fcunit for MaxFC (SI Figure~S2a). Note that these correspond to relative errors of $5.5$\% for MaxFC and $9.2$\% for MinFC, which should ensure accurate predictions of the dimensionalities when combining UMLIPs with FCDimen.

In addition to FCs, we also benchmarked FC-derived quantities, including the phonon frequencies and the FCDimen dimensionality predictions. Phonon dispersions and frequencies had already been benchmarked by several UMLIPs, including MatterSim and MACE, which outperformed other UMLIPs \cite{alexandria}. Here, looking at the maximum phonon frequencies at $\Gamma$-point, both ML models show excellent agreement with DFT, with RMSE of $0.57$ THz for MatterSim (Figure~\ref{fig:benchmark}c) and $0.76$ THz for MACE (SI Figure~S1c).
ML predicted dimensionalities equally reliably. Out of the $4458$ materials screened by the criteria of Ref.~\cite{FCDimen}, MatterSim predicted dimensionalities correctly in $4021$ ($90.2\%$) cases (Figure~\ref{fig:benchmark}d). MACE again performed slightly worse, with $3851$ ($86.4\%$) correct dimensionality predictions (SI Figure~S1d). The agreement between PBE- and PBESol-predicted dimensionalities was $90.0\%$ (SI Figure~S2b)---on par with MatterSim accuracy against PBE. Therefore, based on slightly better performance in predicting materials' FCs and their dimensionalities, we adopt MatterSim as our UMLIP of choice for further calculations.

\subsection{High-throughput calculations}
\begin{figure}[!h]
\begin{center}
  \includegraphics[width=\textwidth]{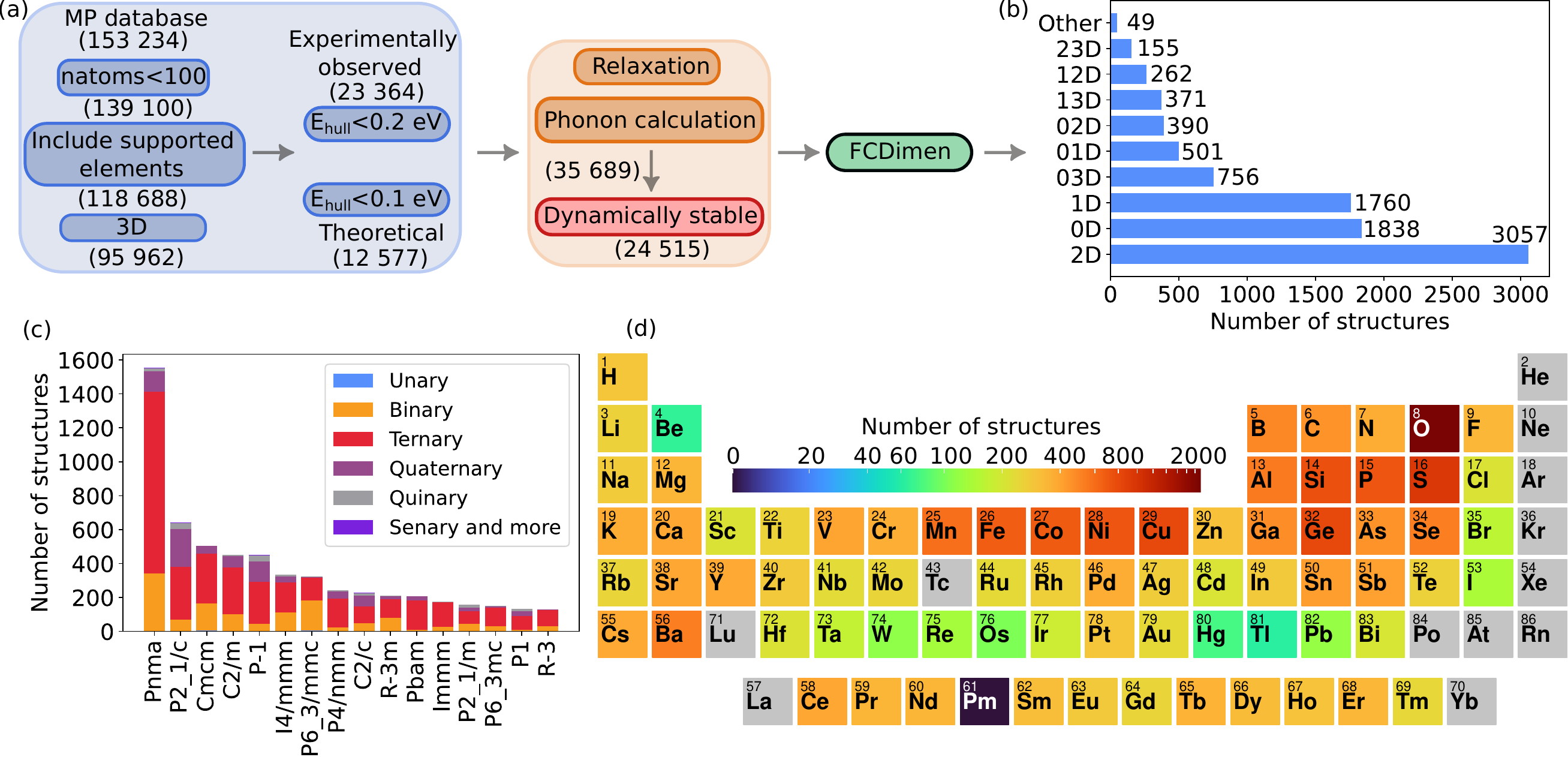}
\end{center}
\caption{\label{fig:results}
High-throughput calculation workflow and resulting dataset statistics.
(a) Workflow for screening and high-throughput calculations. Each step indicates the number of structures involved.
(b) Number of low-dimensional materials discovered in the screened dataset by FCDimen. "Other" corresponds to mixed dimensionalities with more than two types of dimensionalities (012D, 013D, and 023D).
(c) Number of low-dimensional materials classified using the $16$ most common space groups.
(d) Heat map with occurrences of each element in low-dimensional materials in the dataset. 
}
\end{figure}

Armed with a reliable UMLIP, we may now proceed to high-throughput calculations of FCs to discover low-dimensional materials in the Materials Project, an approach previously unfeasible due to the high cost of DFT calculations. The calculations were preceded by screening materials in the MP database (Figure~\ref{fig:results}a). We excluded materials having more than 100 atoms in the unit cell or including elements unsupported by MatterSim, including noble gases, Tc, Lu, Po, At, and Yb. We also excluded materials containing Lu and La due to unreliable results. To direct our search towards stable and undiscovered low-dimensional materials, we further screened the dataset by the following steps: First, we removed materials already previously found to be low-dimensional, as identified by the method of Larsen \etal \cite{Larsen_2019}, as demonstrated in our previous work (Ref.~\cite{FCDimen}), the materials identified as low-dimensional by conventional geometric descriptors are also found to be low-dimensional by FCDimen. Second, we removed materials included in the Phonon database since they had already been studied earlier by DFT and FCDimen \cite{FCDimen}. Third, we removed theoretically predicted materials with energy above the convex hull (E$_{hull}$) larger than $0.1$~eV and experimentally observed materials with E$_{hull}$ larger than $0.2$~eV based on MP data. This screening finally gave us $35{,}689$ materials for the UMLIP-assisted high-throughput phonon calculations.

The phonon calculations gave $24{,}515$ dynamically stable (no imaginary phonon frequencies $>0.1$~THz) materials. Analyzing these materials by FCDimen, we found $9139$ low-dimensional materials including $3057$ sheets (2D), $1760$ chains (1D), $1838$ clusters (0D), and $2484$ mixed dimensionalities (Figure~\ref{fig:results}b). Materials with the highest dimensionality scores are presented in SI Tables S1--S4. 

The discovered low-dimensional materials could be categorized into $16$ most common space groups, with Pnma, P2$_1$/c, and Cmcm being the most common. In addition, binaries, ternaries, and quaternaries were the most commonly appearing species, which can be biased by the nature of the MP database (Figure~\ref{fig:results}c). Most materials contain metalloids (Ge and Si), reactive non-metals (O, S, and P), and transition metals (Fe, Co, Ni) (Figure~\ref{fig:results}d). Promethium has the least number of compounds, and oxides dominate the materials, possibly because most of the materials in MP are oxides. 

\begin{figure}[!h]
\begin{center}
  \includegraphics[width=\textwidth]{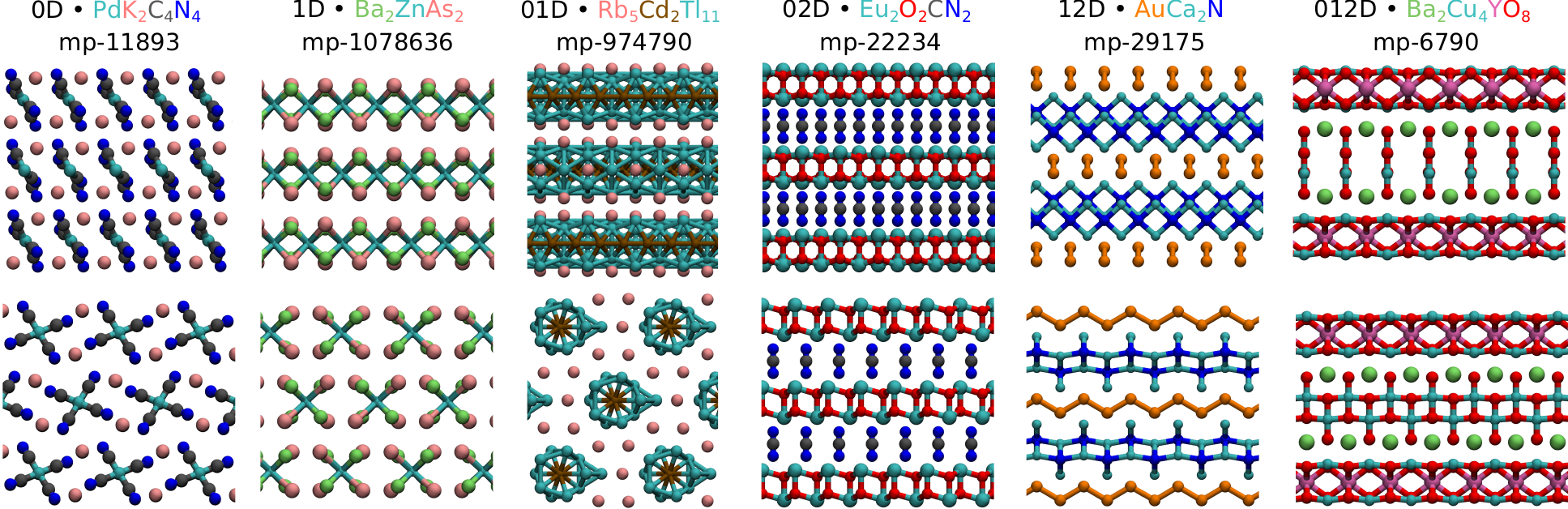}
\end{center}
\caption{\label{fig:other-lowd}
Chemical structures of representative low-dimensional materials in the dataset (top and side views). Legends show dimensionalities, chemical formulas, and MP IDs.
}
\end{figure}

Although low-dimensional units had mostly the same dimensionalities, we also discovered mixed dimensionalities (Figure~\ref{fig:other-lowd}). Note that the geometry-based Larsen algorithm predicted all these materials to be 3D bulk. The 0D units in these materials can take the form of either clusters or isolated atoms. In \ce{Rb2Cd2Tl11}, for example, \ce{Cd2Tl11} nanotubes are surrounded by rubidium atoms. These isolated atoms can be expected to form bulk Rb upon exfoliation. Moreover, in \ce{Eu2O2CN2}, \ce{CN2} molecules are intercalated between \ce{Eu2O2} layers, creating a 02D dimensionality. In \ce{AuCa2N}, 1D \ce{Au} chains are intercalated between 2D \ce{Ca2N} sheets, creating a 12D dimensionality. Here, the behavior of the 1D gold chains during exfoliation is uncertain; they might either be stable or form 2D goldene or bulk gold. A few materials even contain three different dimensionalities. An example of such a material is \ce{Ba2Cu4YO8} (Figure~\ref{fig:other-lowd}), which contains \ce{Cu2YO4} layers, \ce{Cu2O4} chains, and isolated barium atoms. For such a complex material, the products of the 1D and 0D units after exfoliation are unpredictable. 

In addition, we identified $390$ 02D material candidates, $12$ of which have dimensionality scores $\ge0.6$ and a high potential to be etched (SI Table~S5). Finally, we identified $501$ 01D materials, $16$ of them having dimensionality scores $\ge06$ (SI Table~S6).

\subsection{2D materials in detail}
\begin{figure}[!h]
\begin{center}
  \includegraphics[width=\textwidth]{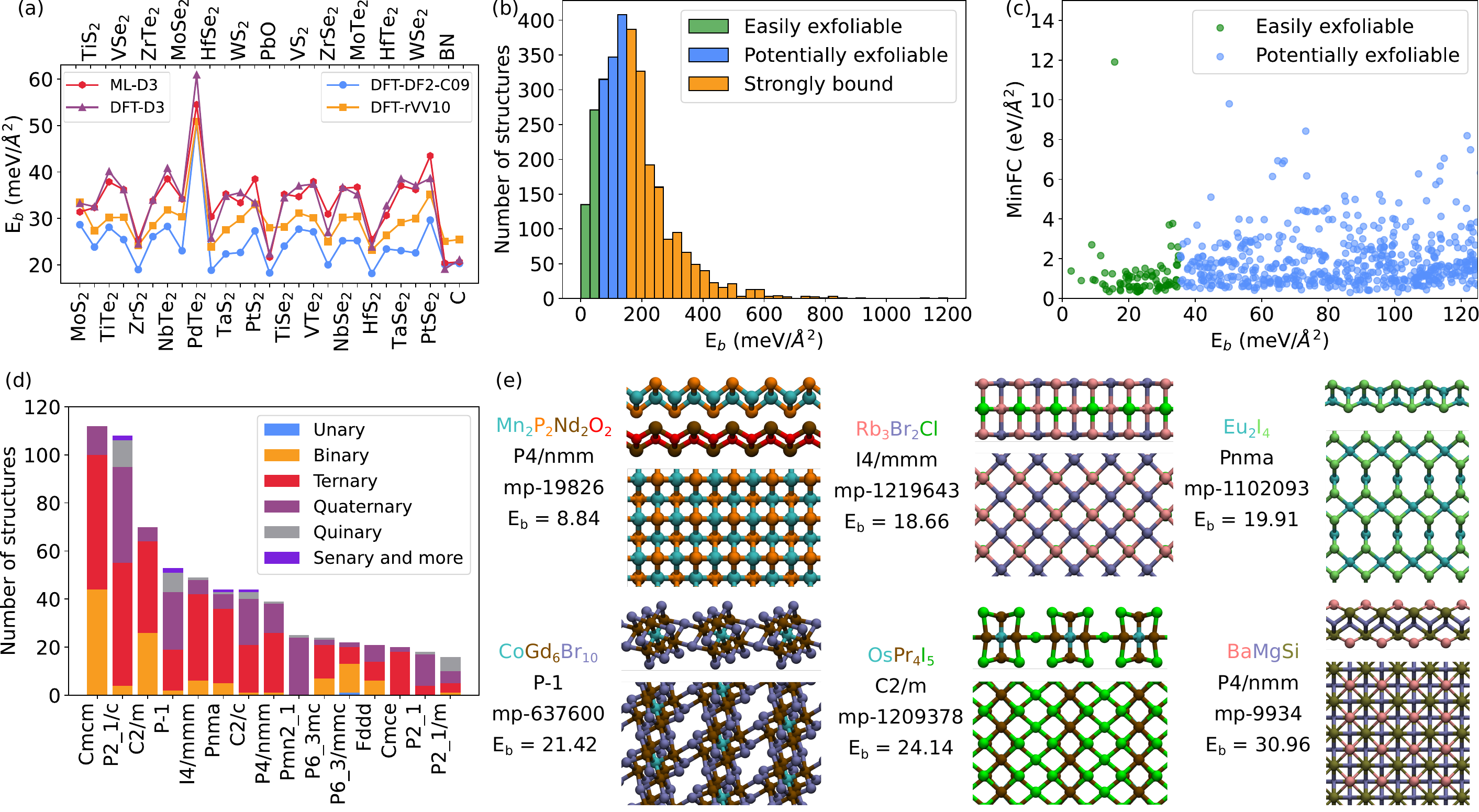}
\end{center}
\caption{\label{fig:d3}
Discovered 2D materials in detail.
  (a) Comparison of the binding energies of some known 2D materials, calculated using ML (with D3–BJ damping) and DFT (DFT-D3, DF2‑C09, and rVV10). DFT-calculated binding energies using DF2‑C09 and rVV10 are taken from Ref.~\cite{Mounet2018}.
  (b) Number of 2D structures as a function of exfoliation energies using ML D3-BJ. Structures are grouped into easily exfoliable ($E_b \leq35$ meV/\AA$^2$), potentially exfoliable ($35<E_b<125$  meV/\AA$^2$), and strongly bound ($E_b>125$ meV/\AA$^2$) 2D materials.
  (c) Correlation between binding energies and MinFC for easily and potentially exfoliable 2D materials.
  (d) Easily and potentially exfoliable 2D materials categorized into $15$ most common space groups.
  (e) Chemical structures (top and side views) of selected easily exfoliable 2D materials from the most common species (binary, ternary, and quaternary). Legends indicate space groups, chemical formulas, MP IDs, and binding energies ($E_b$ in meV/\AA$^2$).
}
\end{figure}

To investigate the discovered 2D materials in more detail, we calculated their interlayer binding energies $E_b$. We accounted for the weak van der Waals interactions by applying the D3 dispersion correction \cite{Grimme2006, Grimme2010, Grimme_BJ, Takamoto2022}. Recently, Sauer \etal~\cite{Kristian2025} demonstrated that UMLIPs with dispersion corrections are ready for large-scale structural simulations of 2D vdW material heterostructures, identifying MatterSim as one of the leading models in their evaluation.
Consistent with these findings, we validated the reliability of ML-D3 calculations by benchmarking our results against DFT-calculated binding energies for a group of known 2D materials (Figure~\ref{fig:d3}a). The binding energies from ML (D3-BJ damping), DFT (D3-BJ, DF2-C09 \cite{Mounet2018}, and rVV10 \cite{Mounet2018}) agree reasonably well. However, most of the D3-BJ binding energies (ML and DFT) are above the two other approaches due to the functional form or damping. We therefore selected our exfoliation energy thresholds slightly higher ($\sim5$ meV/\AA$^2$) than those in Refs.~\cite{Bjorkman2012, Mounet2018, Marzari2023}. 

We grouped the discovered $3057$ 2D materials into three categories based on $E_b$: easily exfoliable ($E_b\leq35$ meV/\AA$^2$), potentially exfoliable ($35<E_b<125$ meV/\AA$^2$), and strongly bound ($E_b>125$ meV/\AA$^2$) (Figure~\ref{fig:d3}b). Consequently, we found $183$ easily exfoliable and $953$ potentially exfoliable 2D materials. Apart from \ce{BN} (mp-2653), all the easily exfoliable materials had MinFC below $5$ \fcunit (Figure~\ref{fig:d3}c). 

The 2D materials are categorized into the $15$ most common space groups, with Cmcm, P2$_1$/c, and C2/m being the most prominent (Figure~\ref{fig:d3}d). Figure~\ref{fig:d3}e shows the chemical structures of the six selected 2D materials from the most common species (binary, ternary, and quaternary). We performed DFT phonon calculations for four non-magnetic 2D materials and compared them with ML-predicted phonon dispersions (SI Figure~S3). Although the agreement for these particular cases is not strong, it validated the accuracy of UMLIPs in identifying true stability, consistent with Ref.~\cite{alexandria}.
ML predicted phonon dispersions of these selected 2D materials are shown in SI Figure~S4. 
We compared the chemical structures, formulas, and Materials IDs (if applicable) of the discovered 2D materials in known databases, including computational two-dimensional database (C2DB) \cite{c2db1, c2db2}, materials cloud two-dimensional structure database (MC2D) \cite{Mounet2018, Marzari2023}, 2DMatPedia \cite{Zhou2019}, 2D topological insulators \cite{topo}, NOMAD (RAE database) \cite{nomad1}, 2D charged building block databases (2DBBs) \cite{Pan2023}, and Materials Project \cite{mp, MP_dimen}. We found $146$ easily exfoliable (SI Table~S7) and $741$ potentially exfoliable 2D materials that could not be found in these databases.

\section{Conclusion}
We demonstrated that both MACE and MatterSim predict interatomic force constants and material dimensionalities with an accuracy comparable to first-principles calculations. Using the slightly better-performing MatterSim, we calculated phonons for $35{,}689$ Materials Project structures, kept the stable ones, and classified their dimensionalities using a method not based on geometries but on true chemical bonding between the atoms. Consequently, we discovered $9139$ low-dimensional materials: $1838$ clusters, $1760$ chains, $3057$ sheets, and $2484$ mixed-dimensional systems. For the discovered 2D materials, we further calculated the interlayer binding energies and found that $146$ are easily exfoliable and $741$ are potentially exfoliable 2D materials. These 2D materials do not exist in the known 2D materials databases. We believe that combining ML with the bond strength-based dimensionality classification method will continue to help discover and explore low-dimensional materials for future applications. We also hope that our discovery inspires experimental work and leads to further expansion of the low-dimensional materials catalog.

\section*{Data availability}
All data supporting the findings have been deposited in the databases at \href{https://doi.org/10.5281/zenodo.17035156}{https://doi.org/10.5281/zenodo.17035156}.

\section*{Supporting information}

The following file is available free of charge.
\begin{itemize}
  \item Supporting Information: Additional details on benchmark results of universal machine-learning interatomic potentials, a list of found low-dimensional materials with the highest score, sample Python scripts to facilitate the usage of databases, and phonon dispersion of selected 2D materials. (PDF)
\end{itemize}

\section*{Acknowledgements}

We acknowledge the Jane and Aatos Erkko Foundation for funding (project EcoMet) and the Research Council of Finland for support under Academy Project No. 357483. The authors thank the EuroHPC Joint Undertaking for awarding this project access to the EuroHPC supercomputer LUMI, hosted by CSC (Finland) and the LUMI consortium through a EuroHPC Regular Access call. We also thank CSC–IT Center for Science Ltd. for generous additional grants of computer time.

\printbibliography

@article{Novoselov2004,
author = {K. S. Novoselov  and A. K. Geim  and S. V. Morozov  and D. Jiang  and Y. Zhang  and S. V. Dubonos  and I. V. Grigorieva  and A. A. Firsov },
title = {Electric Field Effect in Atomically Thin Carbon Films},
journal = {Science},
volume = {306},
number = {5696},
pages = {666-669},
year = {2004},
doi = {https://doi.org/10.1126/science.1102896},
}

@article{Radisavljevic2011,
  doi = {10.1038/nnano.2010.279},
  year = {2011},
  publisher = {Springer Science and Business Media {LLC}},
  volume = {6},
  number = {3},
  pages = {147-150},
  author = {B. Radisavljevic and A. Radenovic and J. Brivio and V. Giacometti and A. Kis},
  title = {Single-Layer {M}o{S}$_2$ Transistors},
  journal = {Nat. Nanotechnol.}
}

@article{Butler2013,
author = {Butler, Sheneve Z. and Hollen, Shawna M. and Cao, Linyou and Cui, Yi and Gupta, Jay A. and Gutiérrez, Humberto R. and Heinz, Tony F. and Hong, Seung Sae and Huang, Jiaxing and Ismach, Ariel F. and Johnston-Halperin, Ezekiel and Kuno, Masaru and Plashnitsa, Vladimir V. and Robinson, Richard D. and Ruoff, Rodney S. and Salahuddin, Sayeef and Shan, Jie and Shi, Li and Spencer, Michael G. and Terrones, Mauricio and Windl, Wolfgang and Goldberger, Joshua E.},
title = {Progress, Challenges, and Opportunities in Two-Dimensional Materials Beyond Graphene},
journal = {ACS Nano},
volume = {7},
number = {4},
pages = {2898-2926},
year = {2013},
doi = {10.1021/nn400280c}
}

@article{Manzeli2017,
  doi = {10.1038/natrevmats.2017.33},
  year = {2017},
  publisher = {Springer Science and Business Media {LLC}},
  volume = {2},
  pages = {17033},
  author = {Sajedeh Manzeli and Dmitry Ovchinnikov and Diego Pasquier and Oleg V. Yazyev and Andras Kis},
  title = {2{D} Transition Metal Dichalcogenides},
  journal = {Nat. Rev. Mater.}
}

@article{Xiang20_Sci,
  author    = {Xiang, Rong and Inoue, Taiki and Zheng, Yongjia and Kumamoto, Akihito and Qian, Yang and Sato, Yuta and Liu, Ming and Tang, Daiming and Gokhale, Devashish and Guo, Jia and Hisama, Kaoru and Yotsumoto, Satoshi and Ogamoto, Tatsuro and Arai, Hayato and Kobayashi, Yu and Zhang, Hao and Hou, Bo and Anisimov, Anton and Maruyama, Mina and Miyata, Yasumitsu and Okada, Susumu and Chiashi, Shohei and Li, Yan and Kong, Jing and Kauppinen, Esko I. and Ikuhara, Yuichi and Suenaga, Kazu and Maruyama, Shigeo},
  title     = {One-Dimensional Van der {W}aals Heterostructures},
  journal   = {Science},
  year      = {2020},
  volume    = {367},
  number    = {6477},
  pages     = {537-542},
  issn      = {0036-8075},
  doi       = {10.1126/science.aaz2570},
  publisher = {American Association for the Advancement of Science}
}

@article{Han19_NComm,
  author  = {Han, Wei and Huang, Pu and Li, Liang and Wang, Fakun and Luo, Peng and Liu, Kailang and Zhou, Xing and Li, Huiqiao and Zhang, Xiuwen and Cui, Yi and Zhai, Tianyou},
  title   = {Two-Dimensional Inorganic Molecular Crystals},
  journal = {Nat. Commun.},
  year    = {2019},
  volume  = {10},
  number  = {1},
  pages   = {4728},
  doi     = {10.1038/s41467-019-12569-9},
}

@article{Zhou_2021,
author = {Zhou, Jin and Bagheri, Mohammad and Järvinen, Topias and Pravda Bartus, Cora and Kukovecz, Akos and Komsa, Hannu-Pekka and Kordas, Krisztian},
title = {{C}$_{60}${B}r$_{24}$/{SWCNT}: A Highly Sensitive Medium to Detect H$_2$S via Inhomogeneous Carrier Doping},
journal = {ACS Appl. Mater. Interfaces},
volume = {13},
number = {49},
pages = {59067-59075},
year = {2021},
doi = {10.1021/acsami.1c16807}
}

@article{Screening2022,
author = {Bagheri, Mohammad and Komsa, Hannu-Pekka},
title = {Screening 0{D} Materials for 2{D} Nanoelectronics Applications},
journal = {Adv. Electron. Mater.},
volume = {9},
number = {1},
pages = {2200393},
doi = {10.1002/aelm.202200393},
year = {2023}
}

@article{Du21_NComm,
	author={Du, Luojun and Zhao, Yanchong and Wu, Linlu and Hu, Xuerong and Yao, Lide and Wang, Yadong and Bai, Xueyin and Dai, Yunyun and Qiao, Jingsi and Uddin, Md Gius and Li, Xiaomei and Lahtinen, Jouko and Bai, Xuedong and Zhang, Guangyu and Ji, Wei and Sun, Zhipei},
	title={Giant Anisotropic Photonics in the 1{D} Van der Waals Semiconductor Fibrous Red Phosphorus},
    journal={Nat. Commun.},
	pages={4822},
	year={2021},
	doi={10.1038/s41467-021-25104-6},
	volume={12},
	number={1},
}

@article{Wei21_AM,
author = {Wei, Nan and Tian, Ying and Liao, Yongping and Komatsu, Natsumi and Gao, Weilu and Lyuleeva-Husemann, Alina and Zhang, Qiang and Hussain, Aqeel and Ding, Er-Xiong and Yao, Fengrui and Halme, Janne and Liu, Kaihui and Kono, Junichiro and Jiang, Hua and Kauppinen, Esko I.},
title = {Colors of Single-wall Carbon Nanotubes},
journal = {Adv. Mater.},
volume = {33},
number = {8},
pages = {2006395},
doi = {10.1002/adma.202006395},
year = {2021}
}

@article{Turunen22_NRevPhys,
  author  = {Turunen, Mikko and Brotons-Gisbert, Mauro and Dai, Yunyun and Wang, Yadong and Scerri, Eleanor and Bonato, Cristian and Jöns, Klaus D. and Sun, Zhipei and Gerardot, Brian D.},
  title   = {Quantum Photonics with Layered 2{D} Materials},
  journal = {Nat. Rev. Phys.},
  year    = {2022},
  volume  = {4},
  pages   = {219},
  doi     = {10.1038/s42254-021-00408-0}
}

@article{Burch2018,
  doi = {https://doi.org/10.1038/s41586-018-0631-z},
  year = {2018},
  publisher = {Springer Science and Business Media {LLC}},
  volume = {563},
  number = {7729},
  pages = {47-52},
  author = {Kenneth S. Burch and David Mandrus and Je-Geun Park},
  title = {Magnetism in Two-Dimensional Van der {W}aals Materials},
  journal = {Nature}
}

@article{Gibertini2019,
  doi = {https://doi.org/10.1038/s41565-019-0438-6},
  year = {2019},
  publisher = {Springer Science and Business Media {LLC}},
  volume = {14},
  number = {5},
  pages = {408-419},
  author = {M. Gibertini and M. Koperski and A. F. Morpurgo and K. S. Novoselov},
  title = {Magnetic 2{D} Materials and Heterostructures},
  journal = {Nat. Nanotechnol.}
}

@article{Klein23_ACSNano,
author = {Klein, Julian and Pingault, Benjamin and Florian, Matthias and Heißenbüttel, Marie-Christin and Steinhoff, Alexander and Song, Zhigang and Torres, Kierstin and Dirnberger, Florian and Curtis, Jonathan B. and Weile, Mads and Penn, Aubrey and Deilmann, Thorsten and Dana, Rami and Bushati, Rezlind and Quan, Jiamin and Luxa, Jan and Sofer, Zdeněk and Alù, Andrea and Menon, Vinod M. and Wurstbauer, Ursula and Rohlfing, Michael and Narang, Prineha and Lončar, Marko and Ross, Frances M.},
title = {The Bulk Van der Waals Layered Magnet {C}r{SB}r is A Quasi-1{D} Material},
journal = {ACS Nano},
volume = {17},
number = {6},
pages = {5316-5328},
year = {2023},
doi = {10.1021/acsnano.2c07316}
}

@article{Kou2017,
author = {Kou, Liangzhi and Ma, Yandong and Sun, Ziqi and Heine, Thomas and Chen, Changfeng},
title = {Two-Dimensional Topological Insulators: Progress and Prospects},
journal = {J. Phys. Chem. Lett.},
volume = {8},
number = {8},
pages = {1905-1919},
year = {2017},
doi = {https://doi.org/10.1021/acs.jpclett.7b00222}
}

@article{Balandin22_MT,
title = {One-Dimensional Van der {W}aals Quantum Materials},
journal = {Mater. Today},
volume = {55},
pages = {74-91},
year = {2022},
issn = {1369-7021},
doi = {https://doi.org/10.1016/j.mattod.2022.03.015},
author = {Alexander A. Balandin and Fariborz Kargar and Tina T. Salguero and Roger K. Lake},
}

@article{Campi2021,
author = {Campi, Davide and Kumari, Simran and Marzari, Nicola},
title = {Prediction of Phonon-Mediated Superconductivity with High Critical Temperature in the Two-Dimensional Topological Semimetal W$_2$N$_3$},
journal = {Nano Lett.},
volume = {21},
number = {8},
pages = {3435-3442},
year = {2021},
doi = {https://doi.org/10.1021/acs.nanolett.0c05125}
}

@article{Kezilebieke2020,
  doi = {10.1038/s41586-020-2989-y},
  year = {2020},
  publisher = {Springer Science and Business Media {LLC}},
  volume = {588},
  number = {7838},
  pages = {424-428},
  author = {Shawulienu Kezilebieke and Md Nurul Huda and Viliam Va{\v{n}}o and Markus Aapro and Somesh C. Ganguli and Orlando J. Silveira and Szczepan G{\l}odzik and Adam S. Foster and Teemu Ojanen and Peter Liljeroth},
  title = {Topological Superconductivity in A Van der {W}aals Heterostructure},
  journal = {Nature}
}

@article{mx_sens,
author = {Hosseini-Shokouh, Seyed Hossein and Zhou, Jin and Berger, Ethan and Lv, Zhong-Peng and Hong, Xiaodan and Virtanen, Vesa and Kordas, Krisztian and Komsa, Hannu-Pekka},
title = {Highly Selective {H}$_2${S} Gas Sensor Based on {T}i$_3${C}$_2${T}$_x$ {MX}ene–Organic Composites},
journal = {ACS Appl. Mater. Interfaces},
volume = {15},
number = {5},
pages = {7063-7073},
year = {2023},
doi = {10.1021/acsami.2c19883}
}

@article{Kim_2018,
author = {Kim, Seon Joon and Koh, Hyeong-Jun and Ren, Chang E. and Kwon, Ohmin and Maleski, Kathleen and Cho, Soo-Yeon and Anasori, Babak and Kim, Choong-Ki and Choi, Yang-Kyu and Kim, Jihan and Gogotsi, Yury and Jung, Hee-Tae},
title = {Metallic {T}i$_3${C}$_2${T}$_x$ {MX}ene Gas Sensors with Ultrahigh Signal-to-Noise Ratio},
journal = {ACS Nano},
volume = {12},
number = {2},
pages = {986-993},
year = {2018},
doi = {10.1021/acsnano.7b07460}
}

@article{Xie2023,
author ={Xie, Minghao and Tang, Sishuang and Zhang, Bowen and Yu, Guihua},
title  ={Metallene-Related Materials for Electrocatalysis and Energy Conversion},
journal  ={Mater. Horiz.},
year  ={2023},
volume  ={10},
issue  ={2},
pages  ={407--431},
publisher  ={The Royal Society of Chemistry},
doi  ={10.1039/D2MH01213H}
}

@article{Zhang_2025,
author = {Zhang, Mo and Wang, Zifeng and Bo, Xin and Huang, Rui and Deng, Dehui},
title = {Two-Dimensional Catalysts: From Model to Reality},
journal = {Angew. Chem., Int. Ed.},
volume = {64},
number = {5},
pages = {e202419661},
doi = {https://doi.org/10.1002/anie.202419661},
year = {2025}
}

@article{Deng2016,
  title = {Catalysis with Two-Dimensional Materials and Their Heterostructures},
  volume = {11},
  ISSN = {1748-3395},
  DOI = {10.1038/nnano.2015.340},
  number = {3},
  journal = {Nat. Nanotechnol.},
  publisher = {Springer Science and Business Media LLC},
  author = {Deng,  Dehui and Novoselov,  K. S. and Fu,  Qiang and Zheng,  Nanfeng and Tian,  Zhongqun and Bao,  Xinhe},
  year = {2016},
  pages = {218–230}
}

@article{D4NR00147H,
author ={Cho, Yun Seong and Kang, Joohoon},
title  ={Two-Dimensional Materials as Catalysts{,} Interfaces{,} and Electrodes for An Efficient Hydrogen Evolution Reaction},
journal  ={Nanoscale},
year  ={2024},
volume  ={16},
issue  ={8},
pages  ={3936-3950},
publisher  ={The Royal Society of Chemistry},
doi  ={10.1039/D4NR00147H}
}

@article{Togo2015,
title = {First Principles Phonon Calculations in Materials Science},
journal = {Scr. Mater.},
volume = {108},
pages = {1-5},
year = {2015},
doi = {https://doi.org/10.1016/j.scriptamat.2015.07.021},
author = {Atsushi Togo and Isao Tanaka}
}

@article{Larsen_2019,
  title = {Definition of A Scoring Parameter to Identify Low-Dimensional Materials Components},
  author = {Larsen, Peter Mahler and Pandey, Mohnish and Strange, Mikkel and Jacobsen, Karsten Wedel},
  journal = {Phys. Rev. Mater.},
  volume = {3},
  issue = {3},
  pages = {034003},
  numpages = {11},
  year = {2019},
  publisher = {American Physical Society},
  doi = {https://doi.org/10.1103/PhysRevMaterials.3.034003}
}

@article{MP_dimen,
  doi = {https://doi.org/10.1557/mrc.2019.94},
  year = {2019},
  publisher = {Springer Science and Business Media {LLC}},
  volume = {9},
  number = {3},
  pages = {874--881},
  author = {Alex M. Ganose and Anubhav Jain},
  title = {Robocrystallographer: Automated Crystal Structure Text Descriptions and Analysis},
  journal = {MRS Commun.}
}

@article{Cheon2017,
author = {Cheon, Gowoon and Duerloo, Karel-Alexander N. and Sendek, Austin D. and Porter, Chase and Chen, Yuan and Reed, Evan J.},
title = {Data Mining for New Two- and One-Dimensional Weakly Bonded Solids and Lattice-Commensurate Heterostructures},
journal = {Nano Lett.},
volume = {17},
number = {3},
pages = {1915-1923},
year = {2017},
doi = {https://doi.org/10.1021/acs.nanolett.6b05229}
}

@article{Gorai2016,
author ={Gorai, Prashun and Toberer, Eric S. and Stevanović, Vladan},
title  ={Computational Identification of Promising Thermoelectric Materials Among Known Quasi-2{D} Binary Compounds},
journal  ={J. Mater. Chem. A},
year  ={2016},
volume  ={4},
issue  ={28},
pages  ={11110-11116},
publisher  ={The Royal Society of Chemistry},
doi  ={https://doi.org/10.1039/C6TA04121C}
}

@article{FCDimen,
author = {Bagheri, Mohammad and Berger, Ethan and Komsa, Hannu-Pekka},
title = {Identification of Material Dimensionality Based on Force Constant Analysis},
journal = {J. Phys. Chem. Lett.},
volume = {14},
number = {35},
pages = {7840-7847},
year = {2023},
doi = {10.1021/acs.jpclett.3c01635}
}

@article{Arkady2022,
author = {Friedrich, Rico and Ghorbani-Asl, Mahdi and Curtarolo, Stefano and Krasheninnikov, Arkady V.},
title = {Data-driven Quest for Two-Dimensional Non-Van der {W}aals Materials},
journal = {Nano Lett.},
volume = {22},
number = {3},
pages = {989-997},
year = {2022},
doi = {https://doi.org/10.1021/acs.nanolett.1c03841}
}

@inproceedings{mace1,
  title={{MACE}: Higher Order Equivariant Message Passing Neural Networks for Fast and Accurate Force Fields},
  author={Ilyes Batatia and David Peter Kovacs and Gregor N. C. Simm and Christoph Ortner and Gabor Csanyi},
  booktitle={Advances in Neural Information Processing Systems},
  editor={Alice H. Oh and Alekh Agarwal and Danielle Belgrave and Kyunghyun Cho},
  year={2022},
}

@misc{mace2,
  title = {The Design Space of E(3)-Equivariant Atom-Centered Interatomic Potentials},
  author = {Batatia, Ilyes and Batzner, Simon and Kov{\'a}cs, D{\'a}vid P{\'e}ter and Musaelian, Albert and Simm, Gregor N. C. and Drautz, Ralf and Ortner, Christoph and Kozinsky, Boris and Cs{\'a}nyi, G{\'a}bor},
  volume = {7},
  ISSN = {2522-5839},
  DOI = {10.1038/s42256-024-00956-x},
  number = {1},
  journal = {Nat. Mach. Intell.},
  publisher = {Springer Science and Business Media LLC},
  year = {2025},
  pages = {56--67}
}

@article{alexandria,
  title = {Universal Machine Learning Interatomic Potentials Are Ready for Phonons},
  volume = {11},
  issn = {2057-3960},
  doi = {10.1038/s41524-025-01650-1},
  pages = {178},
  journal = {npj Comput. Mater.},
  publisher = {Springer Science and Business Media LLC},
  author = {Loew,  Antoine and Sun,  Dewen and Wang,  Hai-Chen and Botti,  Silvana and Marques,  Miguel A. L.},
  year = {2025}
}

@misc{MatterSim,
      title={MatterSim: A Deep Learning Atomistic Model Across Elements, Temperatures and Pressures}, 
      author={Han Yang and Chenxi Hu and Yichi Zhou and Xixian Liu and Yu Shi and Jielan Li and Guanzhi Li and Zekun Chen and Shuizhou Chen and Claudio Zeni and Matthew Horton and Robert Pinsler and Andrew Fowler and Daniel Zügner and Tian Xie and Jake Smith and Lixin Sun and Qian Wang and Lingyu Kong and Chang Liu and Hongxia Hao and Ziheng Lu},
      year={2024},
      eprint={2405.04967},
      archiveprefix={arXiv},
      url={https://arxiv.org/abs/2405.04967}
}

@article{Lyngby2022,
  title = {Data-Driven Discovery of 2{D} Materials by Deep Generative Models},
  volume = {8},
  issn = {2057-3960},
  doi = {10.1038/s41524-022-00923-3},
  number = {1},
  pages = {232},
  journal = {npj Comput. Mater.},
  publisher = {Springer Science and Business Media LLC},
  author = {Lyngby,  Peder and Thygesen,  Kristian Sommer},
  year = {2022}
}

@article{Bjorkman2012,
  title = {Van der {W}aals Bonding in Layered Compounds from Advanced Density-Functional First-Principles Calculations},
  author = {Bj\"orkman, T. and Gulans, A. and Krasheninnikov, A. V. and Nieminen, R. M.},
  journal = {Phys. Rev. Lett.},
  volume = {108},
  issue = {23},
  pages = {235502},
  numpages = {5},
  year = {2012},
  publisher = {American Physical Society},
  doi = {10.1103/PhysRevLett.108.235502}
}

@article{Mounet2018,
  title = {Two-Dimensional Materials from High-Throughput Computational Exfoliation of Experimentally Known Compounds},
  volume = {13},
  issn = {1748-3395},
  doi = {10.1038/s41565-017-0035-5},
  number = {3},
  journal = {Nat. Nanotechnol.},
  publisher = {Springer Science and Business Media LLC},
  author = {Mounet,  Nicolas and Gibertini,  Marco and Schwaller,  Philippe and Campi,  Davide and Merkys,  Andrius and Marrazzo,  Antimo and Sohier,  Thibault and Castelli,  Ivano Eligio and Cepellotti,  Andrea and Pizzi,  Giovanni and Marzari,  Nicola},
  year = {2018},
  pages = {246–252}
}

@article{Marzari2023,
author = {Campi, Davide and Mounet, Nicolas and Gibertini, Marco and Pizzi, Giovanni and Marzari, Nicola},
title = {Expansion of the Materials Cloud 2{D} Database},
journal = {ACS Nano},
volume = {17},
number = {12},
pages = {11268-11278},
year = {2023},
doi = {10.1021/acsnano.2c11510}
}

@article{Grimme2006,
author = {Grimme, Stefan},
title = {Semiempirical GGA-Type Density Functional Constructed with A Long-Range Dispersion Correction},
journal = {J. Comput. Chem.},
volume = {27},
number = {15},
pages = {1787-1799},
doi = {https://doi.org/10.1002/jcc.20495},
year = {2006}
}

@article{Grimme2010,
    author = {Grimme, Stefan and Antony, Jens and Ehrlich, Stephan and Krieg, Helge},
    title = {A Consistent and Accurate Ab Initio Parametrization of Density Functional Dispersion Correction ({DFT-D}) for the 94 Elements {H}-{P}u},
    journal = {J. Chem. Phys.},
    volume = {132},
    number = {15},
    pages = {154104},
    year = {2010},
    issn = {0021-9606},
    doi = {10.1063/1.3382344}
}

@article{Grimme_BJ,
author = {Grimme, Stefan and Ehrlich, Stephan and Goerigk, Lars},
title = {Effect of the Damping Function in Dispersion Corrected Density Functional Theory},
journal = {J. Comput. Chem.},
volume = {32},
number = {7},
pages = {1456-1465},
doi = {https://doi.org/10.1002/jcc.21759},
year = {2011}
}

@article{Takamoto2022,
  title = {Towards Universal Neural Network Potential for Material Discovery Applicable to Arbitrary Combination of 45 Elements},
  volume = {13},
  ISSN = {2041-1723},
  DOI = {10.1038/s41467-022-30687-9},
  number = {1},
  journal = {Nat. Commun.},
  publisher = {Springer Science and Business Media LLC},
  author = {Takamoto,  So and Shinagawa,  Chikashi and Motoki,  Daisuke and Nakago,  Kosuke and Li,  Wenwen and Kurata,  Iori and Watanabe,  Taku and Yayama,  Yoshihiro and Iriguchi,  Hiroki and Asano,  Yusuke and Onodera,  Tasuku and Ishii,  Takafumi and Kudo,  Takao and Ono,  Hideki and Sawada,  Ryohto and Ishitani,  Ryuichiro and Ong,  Marc and Yamaguchi,  Taiki and Kataoka,  Toshiki and Hayashi,  Akihide and Charoenphakdee,  Nontawat and Ibuka,  Takeshi},
  year = {2022}
}

@article{mp,
author = {Jain, Anubhav and Ong, Shyue Ping and Hautier, Geoffroy and Chen, Wei and Richards, William Davidson and Dacek, Stephen and Cholia, Shreyas and Gunter, Dan and Skinner, David and Ceder, Gerbrand and Persson, Kristin a.},
doi = {https://doi.org/10.1063/1.4812323},
journal = {APL Mater.},
number = {1},
pages = {011002},
title = {The Materials Project: A Materials Genome Approach to Accelerating Materials Innovation},
volume = {1},
year = {2013}
}

@article{ase-paper,
  author={Ask Hjorth Larsen and Jens Jørgen Mortensen and Jakob Blomqvist and Ivano E Castelli and Rune Christensen and Marcin
Dułak and Jesper Friis and Michael N Groves and Bjørk Hammer and Cory Hargus and Eric D Hermes and Paul C Jennings and Peter
Bjerre Jensen and James Kermode and John R Kitchin and Esben Leonhard Kolsbjerg and Joseph Kubal and Kristen
Kaasbjerg and Steen Lysgaard and Jón Bergmann Maronsson and Tristan Maxson and Thomas Olsen and Lars Pastewka and Andrew
Peterson and Carsten Rostgaard and Jakob Schiøtz and Ole Schütt and Mikkel Strange and Kristian S Thygesen and Tejs
Vegge and Lasse Vilhelmsen and Michael Walter and Zhenhua Zeng and Karsten W Jacobsen},
  title={The Atomic Simulation Environment—A Python Library for Working with Atoms},
  journal={J. Phys.:Condens. Matter.},
  volume={29},
  number={27},
  pages={273002},
  year={2017}
}

@article{m3gnet,
  title = {A Universal Graph Deep Learning Interatomic Potential for the Periodic Table},
  volume = {2},
  ISSN = {2662-8457},
  DOI = {10.1038/s43588-022-00349-3},
  number = {11},
  journal = {Nat. Comput. Sci.},
  publisher = {Springer Science and Business Media LLC},
  author = {Chen,  Chi and Ong,  Shyue Ping},
  year = {2022},
  pages = {718–728}
}

@article{FIRE,
  title = {Structural Relaxation Made Simple},
  author = {Bitzek, Erik and Koskinen, Pekka and G\"ahler, Franz and Moseler, Michael and Gumbsch, Peter},
  journal = {Phys. Rev. Lett.},
  volume = {97},
  issue = {17},
  pages = {170201},
  numpages = {4},
  year = {2006},
  publisher = {American Physical Society},
  doi = {10.1103/PhysRevLett.97.170201}
}

@article{Burger2025,
author = {Berger, Ethan and Bagheri, Mohammad and Komsa, Hannu-Pekka},
title = {Screening of Material Defects using Universal Machine-Learning Interatomic Potentials},
journal = {Small},
volume = {21},
number = {37},
pages = {e03956},
year = {2025},
doi = {https://doi.org/10.1002/smll.202503956}
}

@article{PRX2013,
  title = {Two-Dimensional Materials from Data Filtering and Ab Initio Calculations},
  author = {Leb\`egue, S. and Bj\"orkman, T. and Klintenberg, M. and Nieminen, R. M. and Eriksson, O.},
  journal = {Phys. Rev. X},
  volume = {3},
  issue = {3},
  pages = {031002},
  numpages = {7},
  year = {2013},
  publisher = {American Physical Society},
  doi = {10.1103/PhysRevX.3.031002}
}

@article{Ashton2017,
  title = {Topology-Scaling Identification of Layered Solids and Stable Exfoliated 2{D} Materials},
  author = {Ashton, Michael and Paul, Joshua and Sinnott, Susan B. and Hennig, Richard G.},
  journal = {Phys. Rev. Lett.},
  volume = {118},
  issue = {10},
  pages = {106101},
  numpages = {6},
  year = {2017},
  publisher = {American Physical Society},
  doi = {10.1103/PhysRevLett.118.106101}
}

@article{Hadeel2022,
  title = {Computational Exfoliation of Atomically Thin One-Dimensional Materials With Application to Majorana Bound States},
  author = {Moustafa, Hadeel and Larsen, Peter Mahler and Gjerding, Morten N. and Mortensen, Jens J\o{}rgen and Thygesen, Kristian S. and Jacobsen, Karsten W.},
  journal = {Phys. Rev. Mater.},
  volume = {6},
  issue = {6},
  pages = {064202},
  numpages = {15},
  year = {2022},
  publisher = {American Physical Society},
  doi = {10.1103/PhysRevMaterials.6.064202}
}

@article{c2db1,
doi = {10.1088/2053-1583/aacfc1},
year = {2018},
publisher = {IOP Publishing},
volume = {5},
number = {4},
pages = {042002},
author = {Haastrup, Sten and Strange, Mikkel and Pandey, Mohnish and Deilmann, Thorsten and Schmidt, Per S and Hinsche, Nicki F and Gjerding, Morten N and Torelli, Daniele and Larsen, Peter M and Riis-Jensen, Anders C and Gath, Jakob and Jacobsen, Karsten W and Jørgen Mortensen, Jens and Olsen, Thomas and Thygesen, Kristian S},
title = {The Computational 2{D} Materials Database: High-Throughput Modeling and Discovery of Atomically Thin Crystals},
journal = {2D Mater.}
}

@article{c2db2,
doi = {10.1088/2053-1583/ac1059},
year = {2021},
publisher = {IOP Publishing},
volume = {8},
number = {4},
pages = {044002},
author = {Gjerding, Morten Niklas and Taghizadeh, Alireza and Rasmussen, Asbjørn and Ali, Sajid and Bertoldo, Fabian and Deilmann, Thorsten and Knøsgaard, Nikolaj Rørbæk and Kruse, Mads and Larsen, Ask Hjorth and Manti, Simone and Pedersen, Thomas Garm and Petralanda, Urko and Skovhus, Thorbjørn and Svendsen, Mark Kamper and Mortensen, Jens Jørgen and Olsen, Thomas and Thygesen, Kristian Sommer},
title = {Recent Progress of the Computational 2{D} Materials Database ({C2DB})},
journal = {2D Mater.},
}

@article{Li2024,
  title = {High-Throughput Screening and Machine Learning Classification of Van der Waals Dielectrics for 2D Nanoelectronics},
  volume = {15},
  ISSN = {2041-1723},
  DOI = {10.1038/s41467-024-53864-4},
  pages = {9527},
  journal = {Nat. Commun.},
  publisher = {Springer Science and Business Media LLC},
  author = {Li,  Yuhui and Wan,  Guolin and Zhu,  Yongqian and Yang,  Jingyu and Zhang,  Yan-Fang and Pan,  Jinbo and Du,  Shixuan},
  year = {2024}
}

@article{Zhou2019,
  title = {{2DM}at{P}edia,  An Open Computational Database of Two-Dimensional Materials from Top-Down and Bottom-Up Approaches},
  volume = {6},
  ISSN = {2052-4463},
  DOI = {10.1038/s41597-019-0097-3},
  pages = {86},
  journal = {Sci. Data},
  publisher = {Springer Science and Business Media LLC},
  author = {Zhou,  Jun and Shen,  Lei and Costa,  Miguel Dias and Persson,  Kristin A. and Ong,  Shyue Ping and Huck,  Patrick and Lu,  Yunhao and Ma,  Xiaoyang and Chen,  Yiming and Tang,  Hanmei and Feng,  Yuan Ping},
  year = {2019}
}

@article{topo,
author = {Marrazzo, Antimo and Gibertini, Marco and Campi, Davide and Mounet, Nicolas and Marzari, Nicola},
title = {Relative Abundance of {Z}2 Topological Order in Exfoliable Two-Dimensional Insulators},
journal = {Nano Lett.},
volume = {19},
number = {12},
pages = {8431-8440},
year = {2019},
doi = {10.1021/acs.nanolett.9b02689}
}

@article{Barnowsky2023,
author = {Barnowsky, Tom and Krasheninnikov, Arkady V. and Friedrich, Rico},
title = {A New Group of 2{D} Non-van der {W}aals Materials with Ultra Low Exfoliation Energies},
journal = {Adv. Electron. Mater.},
volume = {9},
number = {4},
pages = {2201112},
doi = {https://doi.org/10.1002/aelm.202201112},
year = {2023}
}

@article{Deringer19_AM,
author = {Deringer, Volker L. and Caro, Miguel A. and Cs\'{a}nyi, G\'{a }bor},
title = {Machine Learning Interatomic Potentials as Emerging Tools for Materials Science},
journal = {Adv. Mater.},
volume = {31},
number = {46},
pages = {1902765},
doi = {https://doi.org/10.1002/adma.201902765},
year = {2019}
}

@article{Behler2021_review,
author = {Behler, J{\"o}rg},
title = {Four Generations of High-Dimensional Neural Network Potentials},
journal = {Chem. Rev.},
volume = {121},
number = {16},
pages = {10037-10072},
year = {2021},
doi = {10.1021/acs.chemrev.0c00868}
}

@article{Riebesell2024matbench,
  title = {A Framework to Evaluate Machine Learning Crystal Stability Predictions},
  volume = {7},
  ISSN = {2522-5839},
  DOI = {10.1038/s42256-025-01055-1},
  number = {6},
  journal = {Nat. Mach. Intell.},
  publisher = {Springer Science and Business Media LLC},
  author = {Riebesell,  Janosh and Goodall,  Rhys E. A. and Benner,  Philipp and Chiang,  Yuan and Deng,  Bowen and Ceder,  Gerbrand and Asta,  Mark and Lee,  Alpha A. and Jain,  Anubhav and Persson,  Kristin A.},
  year = {2025},
  pages = {836–847}
}

@article{Jinnouchi2025_review,
author = {Jinnouchi, Ryosuke and Minami, Saori},
title = {Machine Learning Force Fields in Electrochemistry: From Fundamentals to Applications},
journal = {ACS Nano},
volume = {19},
number = {25},
pages = {22600-22644},
year = {2025},
doi = {10.1021/acsnano.5c05553}
}

@article{spglib,
  author = {Atsushi Togo, Kohei Shinohara and Isao Tanaka},
  title = {Spglib: A Software Library for Crystal Symmetry Search},
  journal = {Sci. Technol. Adv. Mater., Meth.},
  volume = {4},
  number = {1},
  pages = {2384822--2384836},
  year = {2024},
  doi = {10.1080/27660400.2024.2384822}
}

@article{pymatgen,
title = {Python Materials Genomics (pymatgen): A Robust, Open-Source Python Library for Materials Analysis},
journal = {Comput. Mater. Sci.},
volume = {68},
pages = {314-319},
year = {2013},
doi = {https://doi.org/10.1016/j.commatsci.2012.10.028},
author = {Shyue Ping Ong and William Davidson Richards and Anubhav Jain and Geoffroy Hautier and Michael Kocher and Shreyas Cholia and Dan Gunter and Vincent L. Chevrier and Kristin A. Persson and Gerbrand Ceder},
}

@article{phonopy1,
  author  = {Togo, Atsushi and Chaput, Laurent and Tadano, Terumasa and Tanaka, Isao},
  title   = {Implementation Strategies in Phonopy and Phono3py},
  journal = {J. Phys. Condens. Matter},
  volume  = {35},
  number  = {35},
  pages   = {353001},
  year    = {2023},
  doi     = {10.1088/1361-648X/acd831}
}

\end{document}